\documentclass[aps,prl,reprint,superscriptaddress]{revtex4-2}

\usepackage{graphicx} 
\usepackage{physics}
\usepackage{xcolor}
\usepackage{multirow}
\usepackage{cleveref}
\usepackage{placeins}

\usepackage{amsmath,amsfonts,amssymb,stackengine,graphicx}
\usepackage{mathtools}
\usepackage{ulem}
\usepackage{cancel}

\renewcommand{\cref}[1]{Fig.~\ref{#1}}

\newcommand{\pcf}{\textrm{pCf}}
\newcommand{\ppf}{\textrm{pPf}}

\newcommand{\D}{\vb{D}}
\renewcommand{\S}{\vb \Sigma}

\renewcommand{\u}{\vb{u}}
\newcommand{\change}[1]{{\color{black}{#1}}} 
\newcommand{\delete}[1]{}
\graphicspath{ {./images/} }
\begin{document}

\title{Elastic turbulence in highly entangled polymers and wormlike micelles}

\author{Theo A. Lewy}
\email{tal43@cam.ac.uk}
\affiliation{DAMTP, University of Cambridge, Cambridge CB3 0WA, UK}
%
%
\author{Suzanne M. Fielding}
\affiliation{Department of Physics, Durham University, Science Laboratories, South Road, Durham DH1 3LE, UK}
\author{Peter D. Olmsted}
\affiliation{Department of Physics and Institute for Soft Matter Synthesis \& Metrology, Georgetown University, Washington DC 20057, USA}
\author{Rich R. Kerswell}
\affiliation{DAMTP, University of Cambridge, Cambridge CB3 0WA, UK}

\begin{abstract}
We show theoretically that an initially homogeneous planar Couette flow of a concentrated polymeric fluid is linearly unstable to the growth of two-dimensional (2D) perturbations, within two widely used constitutive models: the Johnson-Segalman model and the Rolie-Poly model. We perform direct nonlinear simulations of both models in 2D to show that this instability leads to a state of elastic turbulence comprising several narrow shear bands that dynamically coalesce, split and interact. Importantly, we show that this 2D instability arises not only in fluids that have a non-monotonic constitutive curve, and therefore show shear banding in 1D calculations, but also in shear thinning fluids with a monotonic constitutive curve, for which an initially homogeneous base state is stable in 1D. For the former category, the high shear branch of the constitutive curve is unstable to 2D instability in both models, so that the high shear band may be turbulent. In the Rolie-Poly model, the low shear branch is also likewise unstable. Our work provides the first simulation evidence for elastic turbulence in highly entangled polymeric fluids. It also potentially explains rheo-chaotic states seen experimentally in shear banding wormlike micelles. We additionally demonstrate elastic turbulence within both models in the planar Poiseuille geometry.
\end{abstract}


\maketitle


Polymeric fluids show strongly nonlinear flow phenomena that significantly impact industrial processes such as the extrusion and moulding of molten plastics~\cite{larson1992instabilities}. At a microscopic level, chainlike polymer molecules show sluggish viscoelastic relaxation dynamics and are easily disturbed by an imposed flow. Beyond a critical value of the Weissenberg number, defined as the ratio of flow rate to viscoelastic relaxation rate, this leads to nonlinear and unstable flow phenomena at the macroscopic level, even at negligible Reynolds number. 

Dilute polymer solutions  have long been known to suffer linear instabilities driven by hoop stresses in flow geometries with curved streamlines~\cite{larson1990purely,olagunju1995elastic,pakdel1996elastic}, with a pathway to elastic turbulence (ET) at high Weissenberg number~\cite{schiamberg2006transitional,Groisman2000}. ET has also been observed in dilute solutions in rectilinear geometries: experimentally in channels~\cite{Pan2013,Qin2017,Shnapp2022} and pipes~\cite{Bonn2011}, and numerically in Kolmogorov flow~\cite{Berti2008, Berti2010, Lewy_Kerswell_2025},
channels~\cite{Rota2024, Lellep2024, Beneitez2024} and planar Couette flow~\cite{Beneitez2023}. 
The instabilities that underlie ET in dilute solutions are predicted to arise either linearly via a hoop stress~\cite{larson1990purely}, centre-mode~\cite{Garg2018,Chaudhary_Garg_Subramanian_Shankar_2021,Khalid_Chaudhary_Garg_Shankar_Subramanian_2021,Khalid2021b} or polymer diffusive~\cite{Beneitez2023, Lewy2024, Couchman2024} mechanism, or via finite-amplitude perturbations \cite{Morozov_2022, Buza2022}. \change{For reviews on these viscoelastic instabilities and ET see \cite{Steinberg2021,Datta2022,Castillo2022}.}

In dilute solutions the chainlike molecules are assumed to behave independently, but in concentrated solutions and melts, they can become highly entangled with each other. Surprisingly, ET in concentrated systems remains significantly under-explored to date. The central contribution of this Letter is to demonstrate ET for the first time in direct numerical simulations  of concentrated shear thinning polymeric fluids. We study two widely used constitutive models in two rectilinear geometries, which lack hoop stresses in any initially laminar base state: planar Poiseuille flow ($\ppf$), in which the base state has significant stress gradients, and planar Couette flow ($\pcf$), in which the base state has uniform stress. We also identify a route to ET via the linear instability of an initially rectilinear base state. In $\ppf$, this is consistent with existing predictions that normal stress gradients can cause instability in shear thinning fluids~\cite{wilson1999instability,wilson2015linear,Barlow2019, grillet2002stability}, as seen experimentally~\cite{picaut2017experimental,bodiguel2015flow,poole2016elastic}. Early work also predicted linear instability of the shear thinning Phan-Thien-Tanner model in $\pcf$~\cite{grillet2002stability}, but without  making any link with ET. Indeed, our finding of ET in concentrated polymeric fluids in a flow geometry that lacks either hoop stresses or stress gradients is particularly unexpected.

In concentrated polymers~\cite{cao2012shear,mohagheghi2015molecular} and wormlike micellar surfactants~\cite{spenley1993nonlinear}, an initially homogeneous shear flow is often unstable to the formation of coexisting bands of differing shear rate, with layer normals in the flow-gradient direction. This instability is driven by a region of negative slope in the underlying constitutive relation between shear stress and shear rate, such that the constitutive curve as a whole is non-monotonic~\cite{YERUSHALMI19701891}. In addition to our work demonstrating ET in fluids with a monotonic constitutive curve, a significant additional contribution of this Letter is to uncover ET in shear banding fluids with a non-monotonic constitutive curve, again by means of both linear stability analysis and nonlinear simulations.

Indeed, experiments on shear banded flows commonly reveal  ET~\cite{Fardin2010, Fardin2012a, Fardin2012b}, with the high shear band often turbid~\cite{BPD97,Lerouge.Decruppe.ea00,becu2007etd,Capp+97b,LDH98,rassolov2022kinetics} and chaotic~\cite{chaos2000,ganapathy2006irr,ghadai2023origin}. Historically, however, shear banded states have often been interpreted via one-dimensional (1D) calculations predicting a flat interface between stationary homogeneous bands~\cite{catesfielding06,olmstedbanding08,Olmsted2000}. Early attempts to capture complex spatiotemporal dynamics suggested possible `rheo-chaos' of the high shear band~{\cite{fielding04,mandal2019complex}}, concentration  coupling~\cite{fielding03a}, or an undulatory instability of the banding interface
~\cite{Fielding,Fielding2010,Wilson_and_Fielding}. Importantly, our work suggests that many experiments should instead be interpreted in terms of the coexistence of an elastically turbulent high shear band  with a laminar low shear band. 

\change{The  ET reported here in concentrated polymeric fluids adds to the list of known systems, including dilute polymeric fluids (as discussed above), active fluids~\cite{Alert2022}, and bubbly air-water flow~\cite{Lance1991}, in which nonlinearities other than momentum advection can produce turbulence.}


\paragraph{Models --- }\label{models}  

We consider incompressible flow that satisfies $\nabla \cdot \u = 0$, and the inertialess force balance equation:
\begin{equation}
\begin{gathered}\label{gov1:momentum}
     0 =  \mu \nabla^2 \vb{u} + \nabla \cdot \S - \nabla p - G_0 \vb{\hat x},
\end{gathered}
\end{equation}
with velocity $\vb{u}$,   solvent viscosity $\mu$, polymer stress $\S$ and  pressure $p$. The imposed pressure gradient  {$G_0\neq 0$} in $\ppf$ and {$G_0=0$} in $\pcf$. For the polymer stress, we consider two widely used constitutive models of entangled polymeric fluids: the phenomenological Johnson-Segalman (JS) model~\cite{Johnson1977} and the microscopically motivated Rolie-Poly  (RP) model~\cite{Likhtman2003}.
Both capture shear thinning and, in some parameter regimes, shear banding and can be written in the same compact way:
\begin{equation}
\begin{gathered}\label{gov3:constitutive}
    \overset{\diamond}{\S} = 2G_{\rm p}\D - \frac{2}{\tau_{\rm R}}(1-A)(\vb I + \S + \beta A \S ) - \frac{\S}{\tau_{\rm d}} + \mathcal{D} \nabla ^2 \S,
\end{gathered}
\end{equation}
with the Gordon-Schowalter time derivative \cite{Gordon72}
\begin{equation}
\begin{gathered}\nonumber
    \overset{\diamond}{\S} \coloneqq \frac{\partial{\S}}{\partial t} + \u \cdot \nabla \S - a(\S \cdot \D + \D  \cdot \S) - (\S  \cdot \vb\Omega - \vb\Omega \cdot \S).
\end{gathered}
\end{equation}
Here $\D$ and $\vb\Omega$ denote the symmetric and antisymmetric parts of the velocity gradient tensor $(\nabla \u)_{ij} = \partial_iu_j$. Both models have a polymer modulus $G_{\rm p}$, reptation time $\tau_{\rm d}$ and stress diffusion coefficient $\mathcal{D}$.  The JS model obtains for $\tau_{\rm R}\to\infty$, with $\beta$ then irrelevant, and has a slip parameter $a$ with $|a| < 1$. The RP model  has a constraint release parameter $\beta$ and Rouse time $\tau_{\rm R}$, with $A  \coloneqq (1 + \tr(\S)/3)^{-1/2}$, and obtains for $a=1$. {For further model details see section I of the Supplementary Material \cite{supp}  (see also references \cite{Burns2020, Wang2008} therein)}.

\paragraph{Flow geometry --- }  We consider two-dimensional (2D) flow in the $x\textrm{-}y$ plane, with flow direction $\vb{\hat x}$ and flow-gradient direction  $\vb{\hat y}$. We assume translational invariance in the vorticity (spanwise) direction $\vb{\hat z}$ ($\partial_z= 0$), with $u_z=0$ and $\Sigma_{xz}=\Sigma_{yz}=0$. Our flow domain is bounded by hard flat plates at $y=\pm h$, where we assume boundary conditions of no-penetration and no-slip for the velocity. In $\pcf$ this gives  $\u|_{\pm h} = \pm U_0 \vb{\hat x}$ where $U_0$ is the magnitude of the plate speed. In $\ppf$ it gives $\u|_{\pm h} = {\bf 0}$. Non-zero stress diffusion $\mathcal{D}$  also requires a boundary condition on $\S$. Unless otherwise stated we use the Neumann condition $\partial_y \S|_{\pm h} = {\bf 0}$. However, we have checked robustness by also considering $\nabla^2 \S|_{\pm h} = {\bf 0}$, equivalent to enforcing  $\mathcal{D}=0$ at the walls~\cite{Sureshkumar1995}.

{\it Units and parameter values --- } We choose units in which the polymer modulus $G_{\rm p}=1$, reptation time $\tau_{\rm d}=1$ and channel half-height $h=1$. We set the dimensionless stress diffusion coefficient $\varepsilon \coloneq \mathcal{D} \tau_d/h^2=10^{-3}$ throughout. In the RP model we also set the constraint release parameter $\beta=0.71$~\cite{Adams2011}. We then vary the slip parameter $a$ in the JS model, the  entanglement number $Z\coloneqq{\tau_d}/{3\tau_R}$ in the RP model, and the dimensionless solvent viscosity ratio $\delta\coloneqq{\mu}/{\tau_d G_{\rm p}}$ in both models in a regime of high polymer concentration, $\delta \ll 1$. We also vary the Weissenberg number $W\coloneqq U_0\tau_d/h$ (in $\pcf$) or pressure gradient $G=G_0h/G_p$ (in $\ppf$).



\begin{figure*}[tb!]
\includegraphics[width=0.9\linewidth]{both_unstable_regions.jpg}  
\caption{Constitutive curves and stability maps in $\pcf$ of the JS model with $\delta=10^{-2}$, $\varepsilon=10^{-3}$ and of the RP model with $\delta=10^{-4}$, $\varepsilon=10^{-3}$.  (a) Constitutive curves of the JS model for several values of $a$. (b) Constitutive curves of the RP model for several values of $Z$. (c) Stability map of the JS model in the $a\textrm{-}W$ plane. (c) Stability map of the RP model in the $Z\textrm{-}W$ plane. Colours show linear instabilities with $0<\omega^*=\omega_{1D}$ (red), $0<\omega_{1D}<\omega^*$ (green) and $\omega_{1D}<0<\omega^*$ (blue).  Hatching for the JS model in c) shows regions where the PDI is the most unstable mode. The PDI does not arise for the RP model in d). The dotted line in d) additionally shows the neutral curve when $\delta=10^{-3}$. Dispersion relations and most unstable eigenmodes at symbols along the horizontal dashed lines in (c) at $a=0.5$ and in (d) at $Z=1000$ are shown in {Fig. S1 in the Supplementary Material}.}
\label{linear_stability1}
\end{figure*}

{\it Linear stability analysis --- } We first examine the stability of an initial base state of homogeneous pCf, with uniform polymer stress $\S$ and shear rate $\dot\gamma$, with respect to small amplitude heterogeneous perturbations $\S', \dot{\gamma}'$ that depend on $x$ and $t$ as $e^{ik(x-ct)}$, as well as on $y$.  
Here $k\in \mathbb{R}$ is the streamwise wavenumber and $c \in \mathbb{C}$ is a {complex wave speed to be found. We define the growth rate $\omega \coloneqq kc_i$, where $c_i\coloneqq\Im c$, so that modes with $\omega>0$ are linearly unstable. The largest growth rate  for purely 1D perturbations (which depend only on $y$) is defined as $\omega_{1D}  \coloneq \omega(k=0)$ while the most unstable growth rate over all $k$ is defined as  $\omega^* \coloneq \sup_{k \in \mathbb{R}} \omega(k)$. Hence $\omega^* \geq \omega_{1D}$, with $\omega^* =\omega_{1D}$ when 1D perturbations are the most unstable. 

Constitutive curves of shear stress as a function of shear rate are shown in Figs.~\ref{linear_stability1}a and~\ref{linear_stability1}b for the JS and RP models respectively. An initial state of homogeneous $\pcf$ (the `0D' state) is known to be unstable to the growth of 1D perturbations in the regime of negative constitutive slope, leading to the formation of shear bands with layer normals along $y$~\cite{YERUSHALMI19701891}. We indeed find $\omega_{1D}>0$ in this regime, in both JS and RP. Interestingly, however, we find purely 1D perturbations to be more unstable than 2D ones  only in the JS model at relatively low shear rates: $0<\omega^*=\omega_{1D}$, as shown by the red lines in Fig.~\ref{linear_stability1}a. Throughout the rest of the regime of negative slope in JS, and all of this regime in RP, 2D perturbations are  more unstable: $0 <\omega_{1D}<\omega^*$, as shown by the green lines in Fig.~\ref{linear_stability1}a,b.  

More interestingly still, in both JS and RP we find that 2D perturbations can grow even in regimes of positive constitutive slope, where 1D perturbations are stable: $\omega_{1D}<0<\omega^*$ (blue  lines in Fig.~\ref{linear_stability1}a,b).  In the JS model, this 2D instability arises only on the high shear branch of the constitutive curve, above the regime of negative constitutive slope. In the RP model, it arises on both the high and low shear branches, either side of the window of negative slope. Importantly, the RP model shows this 2D instability even at low entanglement numbers, for which the  constitutive curve is monotone increasing, giving shear thinning but not shear banding. 

At larger $W$ in the JS model, the 2D instability changes in character, from having an eigenfunction that is global and large scale, corresponding to a bulk constitutive instability,  to being small scale and confined to the boundary, corresponding to a polymer diffusive instability (PDI)~\cite{Beneitez2023, Couchman2024, Lewy2024}). ({These modes are classified in section II of the Supplementary Material \cite{supp}}, and the switch is identified via a discontinuity in the most unstable wavenumber $k^*$ {and a change in the eigenfunctions, see Fig.~S1a \cite{supp}.})  No PDI arises in the RP model: its 2D instability is always a bulk constitutive mode.  

The instability regimes just discussed in the context of the constitutive curves are summarised in the parameter planes of $a\textrm{-}W$ (for JS) and $Z\textrm{-}W$ (RP) in Figs.~\ref{linear_stability1}c,d with the same colour coding as in Figs.~\ref{linear_stability1}a,b. In both models, the large blue regions of parameter space are particularly striking, as they indicate regimes of 2D {\it in}stability in which the constitutive curve nonetheless has positive slope,  giving 1D {\it stability}. We subdivide these blue regions into unhatched, to denote the bulk constitutive instability, and hatched, to denote the PDI (present only in JS).

Shear thinning polymeric fluids were previously shown to be linearly unstable to the growth of 2D perturbations within several constitutive models~\cite{wilson1999instability,Barlow2019,grillet2002stability,wilson2015linear}. However, those works focused mostly on pressure driven flow, $\ppf$, and suggest the instability is driven by gradients in  the base state's normal stress. In significant contrast, we report 2D instability here in the JS and RP models in $\pcf$, for which the base state is homogeneous, consistent with an early report of 2D instability of $\pcf$ in the Phan-Thien-Tanner model of concentrated polymers~\cite{grillet2002stability}.

\paragraph{Elastic Turbulence ---}

\begin{figure}
\includegraphics[width=\linewidth]{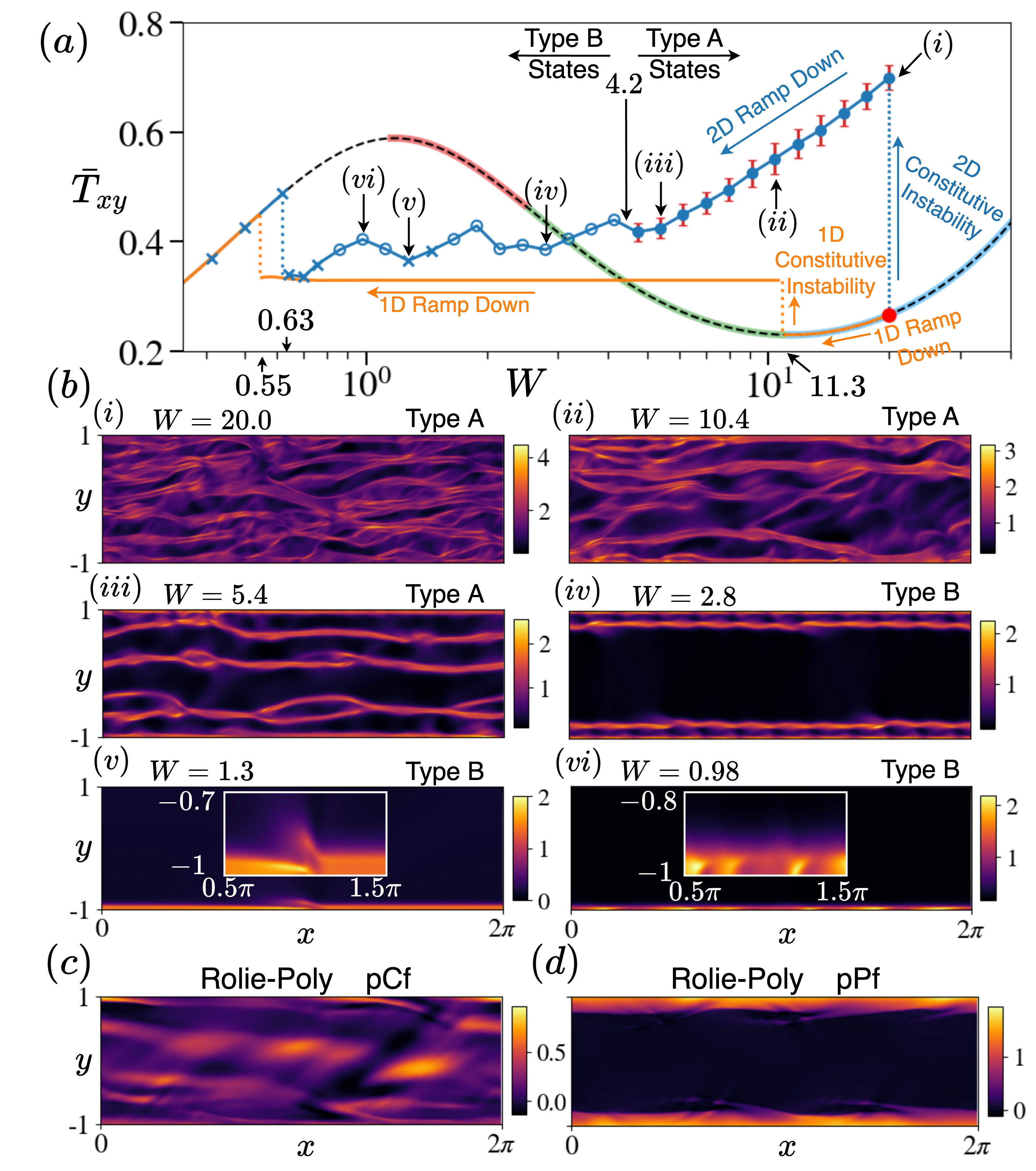} 
\caption{(a) Flow curves obtained by ramping down the shear rate within the JS model in $\pcf$: in 2D (solid blue line), 1D (solid orange line) and 0D (black dashed lines).  In each case the initial condition comprises a state of homogeneous shear at $W=20$ (red circle). In the 2D simulations, steady stress values are shown by crosses and fluctuating stress values by closed (resp. open) circles for standard deviations larger (resp. smaller) then $0.01$, with vertical red bars showing $\pm1$ standard deviation. Linear stability regimes of the 0D curve are marked by colours as in Fig.~\ref{linear_stability1}a. $a=0.5$, $\delta=10^{-2}$, $\varepsilon=10^{-3}$. (b) Snapshots of the trace $\Sigma_{xx}+\Sigma_{yy}+\Sigma_{zz}$ of the polymer stress for states marked $(i)-(vi)$ in (a). (c) Snapshot of the trace of the polymer stress in the RP model in $\pcf$ for $W=20$ and $Z=50$, $\delta=10^{-3}$, $\varepsilon=10^{-3}$. (d) Counterpart  for the RP model in $\ppf$ for $G=-1$. Timeseries and frequency spectra of the stress signals are shown in the Supplementary Material \cite{supp}.}
\label{flow_curve}
\end{figure}

So far, we have shown by means of linear stability analysis that a state of initially homogeneous $\pcf$ suffers a 2D constitutive instability in the large regimes of parameter space shown by the blue unhatched regions in Figs.~\ref{linear_stability1}c,d. We now perform direct nonlinear simulations to evolve this 2D instability to finite amplitude in a domain of length $L_x=2\pi$, using the {implementation described in the Supplementary Material \cite{supp}}.  In both the JS and RP models, this {reveals} chaotic dynamics that we identify as ET. (For justification of our classification of any state as chaotic or quasi-periodic, see \cite{supp}). For comparison, we also perform 1D nonlinear simulations with spatial variations only in $y$.

In both 1D and 2D, we simulate a ramp-down protocol that starts at a high Weissenberg number $W=20.0$, and progressively steps $W$ downwards,  waiting at each $W$ for times $t_w=200$ ($t_w=500$ for $W<5.0$) to ensure that the space-averaged total shear stress $T_{xy}=\Sigma_{xy}+\mu\dot\gamma$ attains a statistically steady state, as evidenced in Fig. S2 of \cite{supp}.  Convergence checks at $W=20.0$ in both models confirmed that numbers of Fourier-Chebyshev modes $(N_x, N_y)=(400,400)$ and $(600,600)$ give results with less than 0.5\% difference in the time- and space-averaged $\bar{T}_{xy}$. For the JS (resp. RP) model we used the latter (resp. former) spatial resolution with timestep $dt=10^{-4}$ (resp. $5\times 10^{-5}$), checking for convergence by halving the timestep.

The flow curve $\bar{T}_{xy}(W)$ obtained in 2D simulations of the JS model is shown by the blue line in  Fig.~\ref{flow_curve}a, with the underlying homogeneous constitutive curve shown by the black dashed line for comparison. Representative snapshots of the states, characterised by the trace of the polymer stress, are shown in (b) for the values of $W$ indicated (i)-(vi) in (a). We identify ET with two  qualitatively different types of state: type A for $W>4.2$ and type B for $W<4.2$. 

Type A states comprise multiple bands arranged across the flow gradient direction, each with a 2D structure. See snapshots $(i)$-$(iii)$ in Fig.~\ref{flow_curve}b. The number of bands fluctuates over time as they dynamically coalesce and split. The associated stress signal  is chaotic, with standard deviation shown by the vertical red bars in Fig.~\ref{flow_curve}a.

Type B states  comprise distinct bands that are localised near the domain walls. Their number remains constant over time, with each band a coherently distinct entity. The shear stress  fluctuates less than for type A states, being either  constant (crosses in Fig.~\ref{flow_curve}a) or with  standard deviation less than $0.01$ (open circles). For $W=2.8$ (snapshot $iv$) we find four bands with quasi-periodic dynamics. At lower $W$ we find a single thin band, still with some 2D structure, exhibiting a traveling wave at $W=1.3$ (snapshot $v$) or chaos at $W=0.98$ ($vi$).  Homogeneous flow is recovered for $W<0.63$.

The flow curve obtained in 1D ramp-down simulations is shown by the orange line for comparison. It follows the well-known scenario of 1D shear banding calculations~\cite{Olmsted2000}. For $W>11.3$ and $W<0.55$ the flow curve coincides with the constitutive curve and the shear field is homogeneous. For $0.55<W<11.3$ the flow curve shows a flat plateau of constant stress $T_{xy}$ and the flow field is shear banded.  (For shear rates just to the left (resp. right) of the negative slope in the constitutive curve, where the stress is greater (resp. less) than the plateau value, states of homogeneous shear are known to be metastable~\cite{grand1997slow}.)

A 1D state with a flat interface between shear bands is known to suffer an interfacial instability~\cite{Fielding, Wilson_and_Fielding}. In a state with just two bands, this leads to undulations along the banding interface~\cite{Fielding2006}. When multiple bands are present (Fig.~5 of Ref.~\cite{Fielding2006}), it leads to a state suggestive of low dimensional chaos. These states reported in Ref.~\cite{Fielding2006} resemble our type B states in Fig.~\ref{flow_curve}b (iv-vi). Plausibly, the 2D constitutive instability of an initial state of homogeneous $\pcf$ reported here takes a system directly to the states obtained via the 2D interfacial instability of an initially 1D shear banded state in Ref.~\cite{Fielding}. No counterpart of our type A states showing broadband turbulence was obtained in Ref.~\cite{Fielding2006}.

So far, we have demonstrated ET across a range of $W$ in the flow geometry of $\pcf$, within the JS model, and with a non-monotonic constitutive curve. We finally find that ET arises robustly across different material parameters (changing $\delta, \varepsilon$), flow geometries ($\pcf$ vs. $\ppf$), constitutive models (JS vs. RP) and shapes of constitutive curve (non-monotonic vs. monotonic). For example, the ET seen in Fig.~\ref{flow_curve}b (i) at $\varepsilon=10^{-3}$, $\delta=10^{-2}$ is also sustained when $\delta$ is reduced to $\delta=2\times10^{-3}$ or increased to $\delta=2\times10^{-2}$ (although not when $\delta=3\times 10^{-2}$). ET is also found if $\varepsilon$ is increased or decreased by a factor of $5$; the bands grow narrower and more fill the domain as $\varepsilon$ falls (not shown). Fig.~\ref{flow_curve}c shows ET in the RP model at parameter values for which the constitutive curve is monotonic. Importantly, this shows that ET can arise in concentrated polymeric fluids that are shear thinning but not susceptible to (1D) shear banding.
We also found ET in $\ppf$,  for which the base state stress $T_{xy}$ is heterogeneous, with the turbulent region  close to the channel walls: for  $G=-1, a=0.5, \delta=10^{-2}, \varepsilon=10^{-3}$ in the JS model (not shown) and in Fig.~\ref{flow_curve}d for the RP model. {Lastly, all ET identified here is 2D, although it no doubt also exists in 3D where simulations at the resolutions needed are currently intractable.}

\paragraph{Discussion---}

In this Letter, we have identified that two widely used constitutive models of concentrated polymeric fluids are susceptible to the 2D instability found for shear thinning fluids in \cite{grillet2002stability}. We have further shown that this triggers 2D ET. 
This is the first numerical demonstration of ET in concentrated polymeric fluids, and it is robust to change in model (JS and RP) and geometry (pCf and pPf). Importantly, ET arises for both monotonic shear thinning and non-monotonic shear banding constitutive curves. In the latter, 2D instability occurs on the high shear branch of both JS and RP models considered, as well as the low shear branch in the RP model. 

\change{Previous studies have reported ET in {\rm dilute} polymeric fluids, arising via hoop stress \cite{Groisman2004}, centre-mode \cite{Lellep2024}, or PDI \cite{Beneitez2023} linear instabilities.  Each of these exists in an Oldroyd-B constitutive model, which is not shear thinning. In contrast, the ET uncovered in this work exists in concentrated polymeric solutions and melts, and with an instability mechanism that stems from the shear thinning behaviour of such fluids. Accordingly, the ET reported here is fundamentally different in mechanism and structure to ET in dilute systems, where shear bands do not seem to play a role.}

Further work is required to see how closely the ET identified here matches experiment. In the Taylor-Couette geometry, constitutive and hoop stress instabilities \cite{Larson1990} coexist, and it is commonly thought that  the latter is responsible for instabilities on the high shear branch \change{of shear-banding wormlike micellar solutions} \cite{Fardin2010, Fardin2012a, Fardin2012b}. However, when the curvature is low and hoop stresses are weak (or absent in rectilinear systems), the constitutive instability is more important and may give rise to a different ET state compared to the hoop stress instability. {Specifically, we expect that curvature-induced instabilities do not exist in the Taylor-Couette geometry provided 
$\sqrt{(r_o-r_i)/r_i} \,\tau_d \dot\gamma<5.92$ \cite{Larson1990} where $r_i, r_o$ are the inner and outer boundary radii, and $\dot\gamma$ is the applied shear, as is consistent with the Pakdel-McKinley criterion \cite{pakdel1996elastic} (the critical value is altered by a factor of $0.7<f<1.3$ during banding \cite{Fardin2011}). The `extremely striated' high shear band in CTAB/NaNO$_3$ in Fig.~12 of Ref. \cite{Lerouge.Decruppe.ea00} ($t=175s$) is comfortably within this regime, with $\sqrt{(r_o-r_i)/r_i} \,\tau_d \dot\gamma = \sqrt{1.5\textrm{mm}/23.5\textrm{mm}}\times 0.17\textrm{s} \times 8\textrm{s}^{-1} = 0.34<5.92$, and hence is a promising setup with which to investigate this constitutive ET. They describe the band as containing `small sub-bands strongly ordered in the direction of the flow' in line with our simulations, but unlike our simulations these sub-bands are localised to the high shear band. While this localisation did not occur in our pCf states, we do obtain localisation by introducing a base shear stress heterogeneity, as is the case in Taylor-Couette (see Fig.~\ref{poiseuille-couette} in the end matter). Other experiments in sheared, hoop stress-free regimes \change{in wormlike micellar solutions} find turbidity and inhomogeneities different to 1D shear bands in CTAB/NaSal \cite{Delgado2009}, and chaotic flow on the high shear branch in EHAC/NH$_4$Cl \cite{Yesilata2006}.} {This instability may also contribute to the so-called `rheo-chaos' observed experimentally by Sood \textit{et al.} \cite{ganapathy2006irr,chaos2000,ganapathy2006trt}, who suggest that concentration-coupling may be required for ET. We have shown this is not necessary.}

Calculations in a shallow 3D Taylor-Couette geometry, likely requiring highly intensive computations, would be a useful next step by allowing for more concrete comparison with experiments. Our findings suggest that the shear banded states observed in wormlike micelles and polymer solutions may comprise a homogeneous low shear rate band co-existing with a turbulent high shear band, in contrast to the simpler 1D interpretation of banding between homogenous stable states \cite{olmstedbanding08}. 


\paragraph{Acknowledgements---} PDO is grateful to Georgetown University and the Ives Foundation for financial support. This project has received funding from the European Research Council (ERC) under the European Union's Horizon 2020 research and innovation programme (grant agreement No. 885146). TAL acknowledges the support of the Cambridge Trust.

\FloatBarrier 

\bibliography{bibliography}


\section{End Matter}\label{end_matter}

\paragraph{Introducing a heterogeneous base shear stress}

Experimentally, the Taylor-Couette geometry is commonly used to study shear banding \cite{Lerouge.Decruppe.ea00,Fardin2010,Fardin2012a, Fardin2012b}, which introduces a heterogeneity into the base shear stress due to its curvature. In contrast, the pCf geometry that was our focus has a homogeneous base shear stress, which results in a degeneracy of the system (in 1D, pCf can support multiple shear banding interfaces, while introducing finite curvature means it can support only one interface \cite{Olmsted2000}). 

Here, we check how including a shear stress heterogeneity into the laminar state can impact ET. This allows for a better comparison with Taylor-Couette experiments, and removes the degeneracy associated with pCf. To do this we take pCf and additionally enforce a pressure gradient with strength $G$, causing the laminar shear stress to vary linearly across the gap, with $dT_{xy}/dy = G$. This results in a Poiseuille-Couette flow, where the flow is driven by both boundary motion and a pressure gradient. Starting from the pCf ET of Fig.~\ref{flow_curve}b (i) at $W=20$, we introduce a pressure gradient of $G=-0.5$, and plot the resulting state in Fig.~\ref{poiseuille-couette}. The ET state is still present, but now the narrow bands are pushed towards the lower boundary, leaving an inactive upper layer. This resembles the Taylor-Couette experiments of \cite{Lerouge.Decruppe.ea00} where curvature was low (see discussion), which similarly introduces heterogeneities into the laminar shear stress.

\begin{figure}[h!]
\includegraphics[width=1\linewidth]{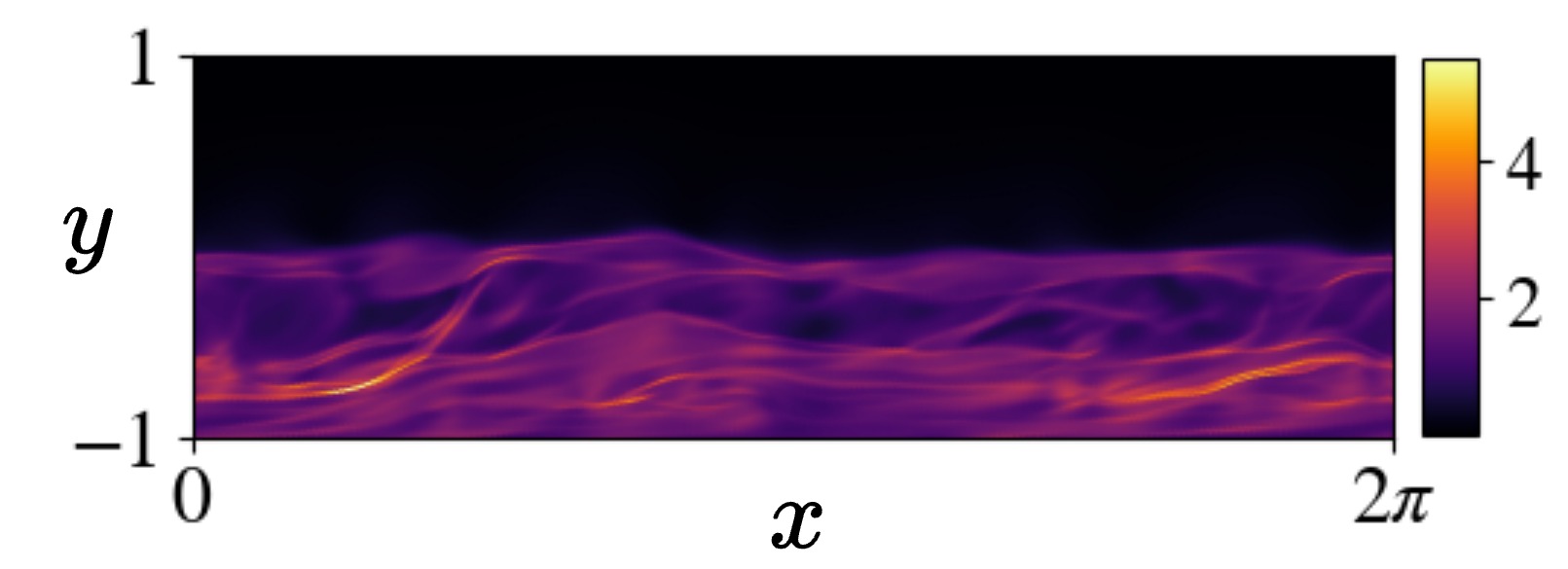} 
\caption{{The trace $\Sigma_{xx}+\Sigma_{yy} +\Sigma_{zz}$ in Poiseuille-Couette flow of the JS model with parameters $G=-0.5$, $W=20$, $\delta=10^{-2}$, $a=0.5$ and $\varepsilon=10^{-3}$.}}
\label{poiseuille-couette}
\end{figure}

\end{document}